\documentclass[11pt,a4paper]{article}

\usepackage[margin=1in]{geometry}
\usepackage{amsmath,amssymb}
\usepackage{graphicx}
\usepackage{booktabs}
\usepackage{siunitx}
\usepackage{xcolor}
\usepackage[hidelinks]{hyperref}
\usepackage{authblk}
\usepackage{caption}
\usepackage{amsthm}
\usepackage{placeins}
\newtheorem{proposition}{Proposition}
\theoremstyle{definition}

\newcommand{\verifynum}[1]{#1}
\newcommand{\rev}[1]{#1}
\newcommand{\revb}[1]{#1}
\newcommand{\revc}[1]{#1}

\newcommand{\TODO}[1]{}

\title{\bfseries How Many Shots Does a Quantum Reservoir Need?\\
Measurement Cost and Its Scaling in Buffered\\ Quantum Reservoir Computing}

\author[1]{Raman Khurana}
\affil[1]{Northwestern University, Evanston, IL, USA}

\date{\today}

\begin{document}
\maketitle

\begin{abstract}
\noindent
Quantum reservoir computing (QRC) inherits from classical reservoir computing the
property that only a linear readout is trained, but unlike a classical reservoir its
state is never directly observable: every feature is an expectation value estimated
from a finite number of measurement shots. The shot budget is therefore a
first-class hyperparameter with no classical analogue. Its effect on QRC has recently
begun to be studied; what remains unsettled is how a required budget should be
defined, and whether competing definitions agree. This paper argues that they do not, and that the
disagreement is systematic rather than incidental.

This paper distinguishes a \emph{precision budget} $N_\varepsilon$, the shots needed to estimate
the reservoir features to a fixed relative accuracy, from a \emph{convergence budget}
$N_\star$, the shots needed for the trained readout to approach its own noise-free
ceiling. Proposition~\ref{prop:confound} shows that $N_\star$ depends on the
noise-free error floor by construction, so that a configuration with a worse floor
appears cheaper, and that configurations should therefore be compared at a fixed
measurement budget rather than by the budget each needs to reach its own ceiling.

Applying this to a small restart-protocol reservoir in simulation, the two budgets
behave differently. $N_\varepsilon$ is invariant to memory horizon (log--log
slope $+0.013$, equivalence confirmed), varies by $0.6\%$ across four tasks whose
achievable error spans a factor of $43$, and rises by under $20\%$ across reservoir
width, where it approaches an upper bound fixed by the tolerance rather than scaling
with size. $N_\star$ varies by three orders of magnitude over the same
sweeps, but Proposition~\ref{prop:confound} shows this cannot be read as measurement
cost: across memory horizon the floor moves by almost $4\times$ in the opposite
direction, which alone accounts for the pattern. Compared at a fixed budget, width
carries a real and steep cost while window length does not.

Further results show that reservoir-seed variance exceeds shot noise by one to two
orders of magnitude at every budget; that buffered memory has a short optimum, attributable to
readout locality rather than to circuit depth; and that two-qubit
$ZZ$ correlators, obtainable from bitstrings already measured, outperformed
single-qubit features at every budget tested.
\end{abstract}

\section{Introduction}
\label{sec:intro}

Reservoir computing (RC) is defined by a fixed, untrained nonlinear dynamical map
followed by a trained linear readout~\cite{jaeger2001,maass2002}. Quantum reservoir
computing (QRC) replaces the classical dynamical system with a quantum
one~\cite{fujii2017,mujal2021}, retaining the property that only the output layer is
optimised. This makes QRC attractive for near-term hardware: no gradients need to be
estimated through the quantum circuit, and the reservoir parameters can be drawn once
and frozen.

QRC differs from its classical counterpart in one respect that has received less
attention than it deserves. A classical reservoir's state is directly readable. A
quantum reservoir's is not: each feature is an expectation value $\langle Z_i\rangle$
that must be estimated from a finite number of projective measurements. The number of
shots therefore sits between the reservoir and the readout as a hyperparameter with
no classical analogue. Too few, and the linear readout is fitting sampling noise; too
many, and measurement time, the dominant cost on real hardware, is spent buying
precision the task never needed.

The consequences of finite sampling in QRC have received increasing attention
(Sec.~\ref{sec:related}), and the qualitative picture, that sampling noise degrades performance and that it interacts with system size, is established. What remains
less settled is quantitative and methodological: how a required budget should
be defined, whether different definitions agree, and how the answer depends on
reservoir width, memory horizon, task and choice of measured observables. Those are
the questions \rev{addressed here}.

\rev{The question is posed deliberately narrowly.} For a small, fixed, hardware-motivated
quantum reservoir on a simple forecasting task, how many shots are actually required
for finite-sampling features to become indistinguishable from their analytic limit,
and how does that requirement scale? \rev{The work runs} entirely in simulation. This is a
deliberate restriction rather than a convenience: only in simulation can the exact
expectation values be computed, which is what allows the sampling contribution to be
isolated cleanly from device noise. The result is a \emph{sampling floor}: a lower bound on the measurement cost of a given QRC configuration, below which no amount of
hardware improvement helps.

\paragraph{Contributions.}
The paper makes four claims, in decreasing order of weight.

\begin{enumerate}\itemsep3pt
\item \textbf{A decomposition that separates two confounded quantities.}
This paper shows that the shot budget at which a trained readout reaches its noise-free
ceiling is, by construction, inversely proportional to that ceiling
(Proposition~\ref{prop:confound}). Configuration comparisons based on this budget are
therefore confounded. \rev{This paper therefore compares configurations at a fixed
measurement budget instead, and shows} empirically that this changes the conclusion:
the apparent variation in required shots across memory horizon is almost entirely an
artefact, whereas the variation across reservoir width is not.

\item \textbf{An empirical characterisation of both budgets} across reservoir width,
memory horizon, task and observable set, with seed ensembles and reported dispersion
(Sec.~\ref{sec:results}). The precision budget admits a closed form that bounds it above and varies
little with either structural parameter; compared at a fixed budget, the sampling penalty grows sharply with width and not with window length.

\item \textbf{Two practical findings.} Reservoir-seed variance exceeds shot noise by
one to two orders of magnitude at every budget, which bears on how QRC results should
be reported. And two-qubit $ZZ$ correlators, which cost no additional circuit
executions because they are computed from bitstrings already measured, dominate
single-qubit features at every budget tested here.

\item \textbf{A resampling procedure} (Sec.~\ref{sec:resampling}) that reproduces shot
statistics exactly at roughly $80\times$ lower cost than repeated shot-based
simulation. This is a convenience rather than a result, but it is what makes
$30$-seed ensembles across ten budgets affordable, and \rev{it is noted} because single-instance reporting is common in this literature,
where cost is plausibly part of the reason.
\end{enumerate}

\section{Related work}
\label{sec:related}

\paragraph{Finite sampling in QRC.}
The effect of finite measurement ensembles on QRC is an active topic. Ahmed
\emph{et al.}~\cite{ahmed2024} analyse finite-sampling noise for chaotic time-series
prediction on Lorenz-63 and a turbulent shear-flow model, showing that noise degrades
conventional QRC more severely than recurrence-free variants because of correlated
noise, and that singular-value truncation of the state matrix improves the
signal-to-noise ratio. Sannia \emph{et al.}~\cite{sannia2025} show that finite-ensemble
noise induces exponential concentration in QRC, which limits resource efficiency as the system is scaled. That result provides a plausible theoretical mechanism for
the width scaling observed here empirically in Sec.~\ref{sec:width}. Nokkala
\emph{et al.}~\cite{nokkala2025} study finite ensembles in continuous-variable QRC and
propose using the measurement cumulative distribution to extract more capacity per
measurement. Kornjaca \emph{et al.}~\cite{rydberg2026} address sampling cost in Rydberg
arrays using randomised and derandomised classical shadows~\cite{huang2020}.
Related work reports that entanglement and coherence can improve resilience to
statistical noise in spin-network reservoirs~\cite{resilience2026}.

This work is complementary rather than competing. Where these studies characterise
how sampling noise degrades performance and propose estimators that mitigate
it, \rev{this paper asks how a required budget should be defined, shows that two
natural definitions scale in opposite directions, and shows that the definition most
often implied (the budget at which the trained readout reaches its own noise-free
ceiling) depends on that ceiling, so that a configuration which performs worse
appears to need fewer shots.}

\paragraph{Measurement cost in variational algorithms.}
The measurement bottleneck is better studied outside RC. Gonthier
\emph{et al.}~\cite{gonthier2022} estimate shot requirements for VQE on molecular
systems and conclude that measurement cost is a principal obstacle to practical
advantage. A substantial body of work reduces that cost through classical
shadows~\cite{huang2020}, locally-biased shadows~\cite{hadfield2022} and adaptive
shot allocation~\cite{kubler2020}. These techniques are not applied here, but it is worth noting in
Sec.~\ref{sec:discussion} that they are the natural route to lowering the floor measured here.

\paragraph{Reservoir characterization.}
The standard tool for characterizing what a reservoir computes is the information
processing capacity \rev{(IPC)} of Dambre \emph{et al.}~\cite{dambre2012}, which
decomposes capacity by memory depth and nonlinearity degree and establishes a
memory--nonlinearity trade-off. It has been applied to quantum reservoirs by
Mart\'inez-Pe\~na \emph{et al.}~\cite{martinezpena2020}. \rev{Task-level mean squared
error is used here instead, because the setting is time-series forecasting and MSE is
the standard figure of merit in that literature; the shot budgets reported below are
therefore directly comparable with the forecasting results they are meant to inform.
The two are complementary rather than alternatives, and
Sec.~\ref{sec:discussion} returns to IPC as a route to characterizing QRC more
fully.}

\section{Setup}
\label{sec:setup}

\subsection{Protocol: buffered (restart) QRC}
\label{sec:protocol}

On gate-based cloud hardware, terminal measurement collapses the register, so a
reservoir cannot in general carry its quantum state from one time step to the next.
Two families of response exist. \emph{Online} or \emph{streaming} protocols use
mid-circuit measurement of a subset of qubits together with deterministic reset, so
that memory persists in the unmeasured subspace; this has been demonstrated on
hardware and shown to process signals of duration not limited by the physical
coherence time~\cite{nisqrc2024,franceschetto2026}. \emph{Buffered} or
\emph{restart} protocols instead rebuild the circuit at each time step from a
classical window of recent inputs, terminating in a full measurement.

\rev{This work adopts} the restart protocol. The reason is access, not principle: mid-circuit and
partial measurement are not exposed through the provider path used for the companion
hardware work, so the streaming family was not reachable. This is a
constraint of the access route and of current devices, not a statement about what is
physically possible; the reader should consult
Refs.~\cite{nisqrc2024,franceschetto2026} for the alternative.

It is worth being precise about what memory the restart protocol has. Inputs are
re-injected sequentially into the same register without reset, so the state
evolves across the window under quantum dynamics and the resulting feature vector
depends on the ordering of the window, not merely its contents. This is genuine
quantum temporal processing, and distinguishes the protocol from a quantum extreme
learning machine in which the whole window is encoded in parallel into a shallow
circuit~\cite{mujal2021}. What is classically imposed is the maximum horizon:
inputs older than the window length were never injected and cannot influence the
features. Within the window, how much of each past input survives, and how the inputs
mix nonlinearly, is set by the reservoir dynamics.

\subsection{Circuit}

The reservoir uses $n_{\mathrm{res}}$ reservoir qubits and one input qubit,
$n_q = n_{\mathrm{res}} + 1$. Reservoir parameters
$\theta \in \mathbb{R}^{n_{\mathrm{res}} \times 2}$ are drawn once from
$\mathcal{U}(0, 2\pi)$ and never updated. For an input window
$(x_{t-k+1},\dots,x_t)$ of length $k$, each element is processed by the block
\begin{enumerate}\itemsep1pt
\item $R_Y(x)$ on the input qubit;
\item $\mathrm{CNOT}$ from the input qubit to each reservoir qubit (star fan-out);
\item $R_Y(\theta_{i,0})\,R_Z(\theta_{i,1})$ on each reservoir qubit $i$;
\item a ring of $\mathrm{CNOT}$s among the reservoir qubits.
\end{enumerate}
The circuit terminates in a measurement of all $n_{\mathrm{res}}$ reservoir qubits;
the input qubit is not measured. Scalar inputs are min--max scaled into $[0,\pi]$
before encoding. The star topology and shallow per-step block were chosen with
hardware execution in mind.

Because each window element contributes its own block and no reset occurs between
blocks, both size parameters enter the circuit cost. For a window of length $k$ the measured cost is exactly
\begin{equation}
\mathrm{depth} = k\,(n_{\mathrm{res}} + 4) + 2,
\qquad
\#\mathrm{CNOT} = 2\,k\,n_{\mathrm{res}},
\label{eq:depth}
\end{equation}
verified over $k \in [1,12]$ and $n_{\mathrm{res}} \in \{2,4,6,8\}$. The depth is
dominated by the star fan-out: the $n_{\mathrm{res}}$ CNOTs in step~2 all share the
input qubit as control and therefore serialise completely, accounting for
$n_{\mathrm{res}}$ of the $n_{\mathrm{res}}+4$ layers per block.

\subsection{Readout and metrics}

Features $X \in \mathbb{R}^{T \times d}$ are the estimated expectation values at each
time step. The target is the driving signal advanced by one step. \rev{The first $W = 10$ steps
are discarded} as washout, since windows are truncated at the start of the series,
and the remainder is split $70/30$ into train and test. The readout is ordinary least squares
with a bias column, except in Appendix~\ref{app:ridge} where ridge regression is used and
the regularisation strength is selected on a held-out validation segment.

\subsection{Tasks}

\rev{This work uses} one-step-ahead prediction on a noisy sinusoid, on the $x$-component of the
Lorenz-63 system, and on the Mackey--Glass delay system, together with the
NARMA-10 benchmark~\cite{atiya2000}. The first three are \emph{autonomous}: the
reservoir is driven by the signal and predicts that same signal one step later.
NARMA-10 is \emph{driven}: \rev{the reservoir sees an exogenous i.i.d.\ input and must
predict a separate series that is autoregressive in its own past and quadratic in
both that past and the drive, with a dependence on inputs nine steps apart. The
reservoir is never shown the target series, so its state has to be reconstructed from
the drive history alone. NARMA-10 is therefore a different kind of problem from the
first three rather than a harder instance of the same one.}

\section{\revb{Exact-distribution sampling}}
\label{sec:resampling}

\rev{Characterising a quantum reservoir across shot budgets requires features at many
budgets, many repeats, many reservoir seeds and many structural configurations.} Simulating each
combination with a shot-based backend is wasteful, because the underlying quantum
state does not depend on the shot count.

\rev{The method here instead computes}, once per input window, the exact measurement distribution
$p \in \mathbb{R}^{2^{n_{\mathrm{res}}}}$ over computational-basis outcomes of the
reservoir register\revb{, with $p_b = |\langle b | \psi\rangle|^2$ and
$\sum_b p_b = 1$}. Estimated expectation values at $N$ shots are then obtained by
drawing a count vector $c \sim \mathrm{Multinomial}(N, p)$ and forming
\begin{equation}
\widehat{\langle Z_i\rangle} = \sum_{b} \frac{c_b}{N}\,s_{b,i},
\qquad
s_{b,i} = 1 - 2\,\mathrm{bit}_i(b),
\end{equation}
\revb{where $c_b$ is the number of shots returning bitstring $b$ and
$\mathrm{bit}_i(b) \in \{0,1\}$ is the value of qubit $i$ in $b$, so that
$s_{b,i} = +1$ for outcome $\lvert 0 \rangle$ and $-1$ for $\lvert 1 \rangle$. The analytic value
is recovered by substituting $p$ for $c/N$.} Two-qubit
correlators follow from the same distribution as
$\widehat{\langle Z_iZ_j\rangle} = \sum_b (c_b/N)\, s_{b,i}s_{b,j}$, requiring no
additional measurement. This is not an approximation: it is exactly the sampling
process a shot-based simulator performs.

\revb{Note that this is sampling from an exactly known distribution, not resampling
from previously drawn shots: $p$ is computed from the statevector rather than
estimated, so the draws carry no error beyond the multinomial noise of the $N$ shots
themselves.} This was verified numerically against a shot-based simulator. Analytic values agree to
machine precision ($\max|\Delta| \approx 10^{-16}$); over $200$ independent draws at
$N = \num{20000}$, both routes reproduce the analytic mean and the theoretical spread
$\sqrt{(1 - \langle Z\rangle^2)/N}$. The \revb{exact-sampling} route was approximately
$80\times$ faster, and vectorising the multinomial draw across time steps contributed
a further factor of two. Every result below rests on this equivalence.

\section{Two notions of ``required shots''}
\label{sec:metrics}

This paper reports two threshold definitions, both computed per reservoir seed and
summarised by the median and interquartile range across seeds, with log-interpolation
between grid points.

\paragraph{\revb{Notation.}}
\revb{Write $X_N \in \mathbb{R}^{T \times d}$ for the feature matrix obtained at a
budget of $N$ shots per window, with one row per retained time step ($T$ rows after
washout) and one column per measured observable ($d = n_{\mathrm{res}}$ for
single-qubit $Z$ readout). Write $X_\infty$ for the same matrix built from exact
expectation values, that is the $N \to \infty$ limit, which simulation gives directly.
Let $\|\cdot\|_F$ denote the Frobenius norm. Let $\mathrm{MSE}(N)$ be the test error of
the readout trained and evaluated on $X_N$, and $\mathrm{MSE}_\infty$ the same quantity
for $X_\infty$. The latter is the \emph{noise-free floor}: the error the configuration
attains when sampling is not a limitation, and the value $\mathrm{MSE}(N)$ approaches
from above as $N$ grows.}

\paragraph{Precision budget $N_\varepsilon$.} The smallest budget $N$ at which the
normalised feature error falls below a tolerance,
\begin{equation}
\varepsilon(N) \;=\; \frac{\|X_N - X_\infty\|_F}{\sqrt{Td}} \;\le\; \tau_{\mathrm{feat}},
\qquad \tau_{\mathrm{feat}} = 0.01 .
\end{equation}
\revb{Dividing by $\sqrt{Td}$ makes $\varepsilon(N)$ the root-mean-square deviation of
a single feature entry, so the tolerance has the same meaning whatever the matrix
dimensions: $\tau_{\mathrm{feat}} = 0.01$ asks that a typical estimated expectation
value sit within $0.01$ of its exact counterpart, on a scale where
$\langle Z_i \rangle \in [-1, 1]$.} This measures estimator precision alone and is
independent of the readout.

\paragraph{Convergence budget $N_\star$.} The smallest budget at which the test error
approaches the noise-free floor,
\begin{equation}
\mathrm{MSE}(N) \;\le\; (1 + \tau_{\mathrm{mse}})\,\mathrm{MSE}_\infty,
\qquad \tau_{\mathrm{mse}} = 0.05 ,
\end{equation}
\revb{that is, the budget at which the readout comes within $5\%$ of the error it would
achieve with perfectly estimated features.} This is the quantity most often meant
implicitly by ``how many shots are needed'', and the one this paper argues against
using for comparison.

\revb{The difficulty is visible in the definition itself. The target
$(1 + \tau_{\mathrm{mse}})\,\mathrm{MSE}_\infty$ is not a fixed level of error: it moves
with $\mathrm{MSE}_\infty$, and so differs from one configuration to the next. A
configuration that performs well sets itself a demanding target, while one that
performs poorly sets itself a lenient one. The following makes that precise.}

\begin{proposition}[Floor confounding]
\label{prop:confound}
\revb{Let $E(N) = \mathrm{MSE}(N) - \mathrm{MSE}_\infty \ge 0$ denote the excess risk,
the part of the test error attributable to finite sampling. Suppose it factorises as}
\begin{equation}
E(N) \;=\; c \; g(N),
\label{eq:factorise}
\end{equation}
\revb{where $g$ is strictly decreasing with $g(N) \to 0$ as $N \to \infty$ and
describes how fast sampling noise is suppressed by the budget, and the constant
$c > 0$ describes how strongly the trained readout converts a given level of feature
noise into predictive error. Then the convergence budget is}
\begin{equation}
N_\star \;=\; g^{-1}\!\left(\frac{\tau_{\mathrm{mse}}\,\mathrm{MSE}_\infty}{c}\right),
\label{eq:nstar}
\end{equation}
and in the particular case $g(N) = 1/N$,
\begin{equation}
N_\star \;=\; \frac{c}{\tau_{\mathrm{mse}}\,\mathrm{MSE}_\infty}.
\label{eq:nstar1}
\end{equation}
\end{proposition}

\begin{proof}
By definition $N_\star$ is the least $N$ satisfying
$\mathrm{MSE}(N) \le (1+\tau_{\mathrm{mse}})\,\mathrm{MSE}_\infty$. Subtracting
$\mathrm{MSE}_\infty$ from both sides, this is equivalent to
$E(N) \le \tau_{\mathrm{mse}}\,\mathrm{MSE}_\infty$. Substituting
\eqref{eq:factorise} gives $g(N) \le \tau_{\mathrm{mse}}\,\mathrm{MSE}_\infty / c$.
\revb{Since $g$ is strictly decreasing it is invertible, and $g^{-1}$ is itself
decreasing, so the least such $N$ is the one at which the inequality becomes an
equality}, which is~\eqref{eq:nstar}. Putting $g(N) = 1/N$, so that
$g^{-1}(y) = 1/y$, gives~\eqref{eq:nstar1}.
\end{proof}

\revb{Two features of~\eqref{eq:nstar} matter. First, $\mathrm{MSE}_\infty$ appears on
the right-hand side, so the budget depends on the floor it is measured against.
Second, $g^{-1}$ is decreasing, so a \emph{smaller} floor produces a \emph{larger}
$N_\star$. Under the special case~\eqref{eq:nstar1} the relation is exactly inverse:
halving the floor doubles the budget, with $c$ unchanged.}

A configuration can therefore appear to need more shots purely because it performs
better, and conversely a poor configuration appears cheap. \revb{$N_\star$ mixes two
quantities a reader will naturally conflate: how sensitive the readout is to sampling
noise, which is $c$, and how good the configuration is, which is
$\mathrm{MSE}_\infty$.} This is not a small correction: \rev{Sec.~\ref{sec:decomposition} reports a
sweep} in which the entire observed variation in $N_\star$ is of this kind.

\revb{It is worth being clear about which part of Proposition~\ref{prop:confound} the
data support. Equation~\eqref{eq:nstar} assumes only that $g$ is strictly decreasing,
so it holds whatever the shape of the decay, and the confounding it establishes is
therefore unaffected by anything measured below. Equation~\eqref{eq:nstar1} is the
narrower statement that follows when $g(N) = 1/N$, and it is that special case alone
which the measurements set aside.}

\revb{Fitting $E(N) \propto N^{-\beta}$ over $N \in [10^2, 3\times10^5]$, that is,
estimating the decay exponent by least squares on $\log E$ against $\log N$, gives
$\beta$ not close to unity in general:}
$\beta = 1.04$ at $n_{\mathrm{res}}=2$, $0.81$ at $n_{\mathrm{res}}=3$, and $0.61$ and
$0.62$ at $n_{\mathrm{res}}=4,5$ (all $R^2 > 0.97$). \revb{The decay is therefore not
$1/N$, and $\beta$ is not even common to the configurations studied, so
\eqref{eq:nstar1} does not apply here.}

\revb{The consequence is narrow but useful. Inverting \eqref{eq:nstar1} invites the
normalisation $c = N_\star\,\tau_{\mathrm{mse}}\,\mathrm{MSE}_\infty$, which looks like
a way to strip the floor out of $N_\star$ and recover a comparable measure of readout
sensitivity. That normalisation is valid only under $g(N) = 1/N$. With $\beta$ varying
the implied coefficient is not constant even within one configuration, and its units
change with $\beta$, so it cannot be compared across configurations at all. This is
worth flagging because it is an easy step to take, and on the data here it yields a
scaling in the opposite direction to the truth. What fails is the inversion, not the
proposition: \eqref{eq:nstar} stands, and with it the conclusion that $N_\star$ should
not be used to rank configurations.}

This paper therefore avoids deriving a coefficient, and \revb{compares} configurations
at a \emph{fixed reference budget} instead. For a stated $N_{\mathrm{ref}}$ \revb{it
reports} the excess risk $E(N_{\mathrm{ref}})$ and the relative sampling penalty
\begin{equation}
\rho(N_{\mathrm{ref}}) \;=\; \frac{E(N_{\mathrm{ref}})}{\mathrm{MSE}_\infty}
\;=\; \frac{\mathrm{MSE}(N_{\mathrm{ref}}) - \mathrm{MSE}_\infty}{\mathrm{MSE}_\infty}.
\label{eq:rho}
\end{equation}
Unlike $N_\star$, $\rho$ does not invert when the floor moves: it asks how much a given
budget costs you, not how much budget a moving target demands.

\revb{Both thresholds are read off a finite grid of budgets, and for some reservoir
seeds the curve never crosses the tolerance anywhere on it. Such a seed yields no
number: the run establishes only that its threshold lies above the largest budget
tested, $3\times10^6$. Borrowing the term from survival analysis, that observation is
\emph{censored}.}

\revb{Censoring is not spread evenly over seeds. A seed fails to cross precisely
because its threshold is large, so the seeds that are lost are systematically the
most expensive ones. A median taken over the seeds that did cross is therefore a
median of the easier subset, and it understates the value a median over all seeds
would take. The understatement grows with the censored fraction, and at the largest
reservoir widths studied that fraction approaches half.}

\revb{Two conventions follow. The fraction of censored seeds is reported alongside
every threshold, so that the reader can judge how much weight the figure carries. And
where that fraction is substantial, the median is read as a lower bound on the budget
rather than as an estimate of it: the true requirement is at least this large, and
how much larger the data do not say.}

\section{Results}
\label{sec:results}

Unless stated otherwise: $T = 200$ time steps, washout $W = 10$, $30$ reservoir
seeds, $20$ shot-noise repeats per seed, and a shot grid spanning
$3\times10^1$ to $3\times10^6$.

\subsection{\revc{Convergence, its spread, and its mechanism}}
\label{sec:convergence}

\begin{figure}[t]\centering
\includegraphics[width=0.325\textwidth]{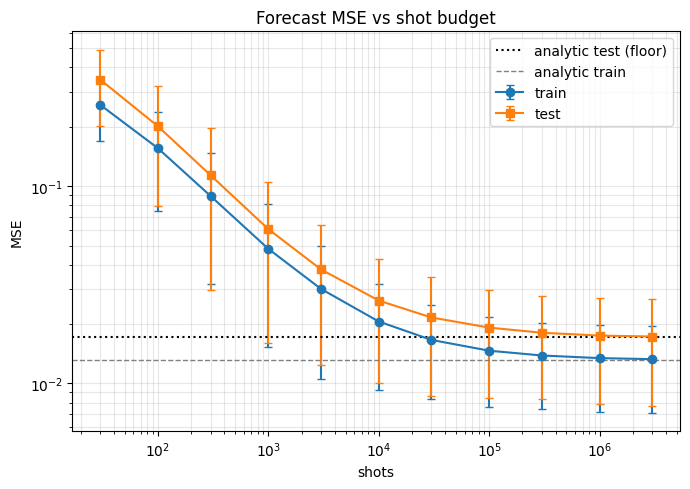}\hfill
\includegraphics[width=0.325\textwidth]{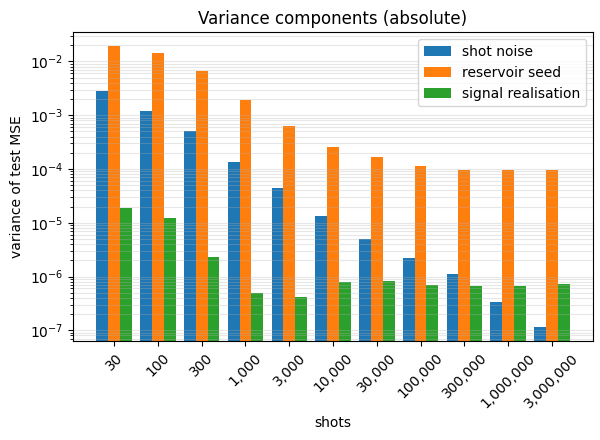}\hfill
\includegraphics[width=0.325\textwidth]{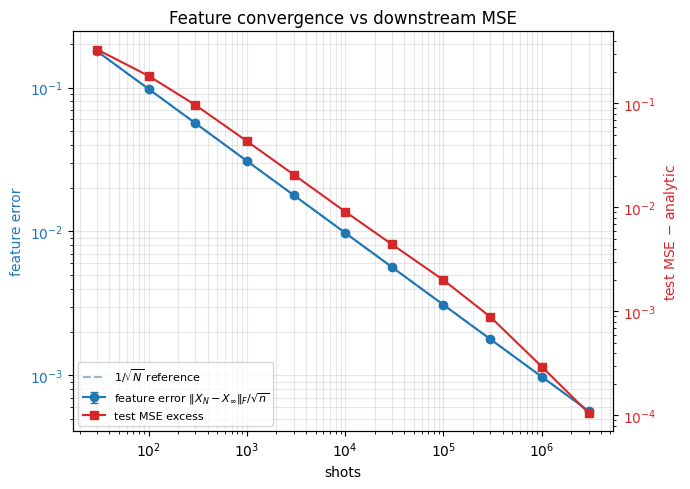}
\caption{\revc{\textbf{Left:} train and test MSE against shot budget, with the
noise-free values as horizontal asymptotes; error bars are a pooled standard
deviation over reservoir seeds, data seeds and shot-noise repeats.
\textbf{Middle:} that spread separated into its three components.
\textbf{Right:} feature error $\varepsilon(N)$ against the $1/\sqrt{N}$ reference
(left axis, blue) and excess risk $E(N)$ over the noise-free floor (right axis, red).}}
\label{fig:convergence}
\end{figure}

\revc{The left panel of Fig.~\ref{fig:convergence} shows test and train error against
shot budget, with the noise-free values each attains from exact features drawn as
horizontal asymptotes. Both curves fall monotonically and approach their asymptotes
strictly from above: sampling noise can only degrade the readout, so
the floor is approached and never crossed. The approach is slow. At $30$ shots the
test error is roughly twenty times its floor; at $10^4$ it remains about $50\%$ above
it; only in the final decade of the grid does it settle. The train curve lies below
the test curve throughout by the usual generalisation margin, and the two floors
differ by that same margin, which confirms the gap belongs to the readout rather than
to sampling.}

\revc{The error bars in that panel are wide at every budget and, in relative terms, do
not narrow as the budget grows. They are a single pooled standard deviation over three
independent sources of randomness, which are worth separating. The \emph{reservoir
seed} draws the fixed rotation angles $\theta$ that define the circuit, so a different
seed is a different untrained dynamical system with the same topology; thirty are
used. The \emph{data seed} generates the series itself, which for the sinusoid means
the observation noise added to the clean signal; three are used. The \emph{shot
repeat} holds reservoir and data fixed and simply measures again, twenty times per
combination. The three answer different questions: whether a different reservoir would
have given the same answer, whether a different sample of the data would, and whether
a rerun of the identical experiment would.}

\revc{The middle panel decomposes the variance of the test error into exactly those
three components. Reservoir-seed variance dominates shot noise at every budget, by a
factor of} \verifynum{$\sim\!11\times$} \revc{at $10^2$ shots widening to}
\verifynum{$\sim\!27\times$} \revc{at $10^5$, and signal realisation is negligible
beside both. Because the seed contribution does not depend on the shot budget, it is
most of what the left panel's error bars are showing, and it is why they fail to
narrow: by the right-hand end they reflect how much the noise-free floor itself varies
between reservoir draws rather than any residual sampling noise. Appendix~\ref{app:variance}
gives the same decomposition as a share of the total.}

This has a direct methodological consequence. A QRC result reported for a single
reservoir instance is dominated by a variance component that has nothing to do with
the quantum dynamics being studied. Without seed ensembles, differences
between configurations are not resolvable.

\revc{The right panel separates the two quantities whose interaction produces the
left. Its left axis, in blue, carries the feature error $\varepsilon(N)$, measuring how
far the estimated expectation values lie from exact; its right axis, in red, carries
the excess risk $E(N)$. The $1/\sqrt{N}$ reference is invisible because the blue curve
lies on top of it: the estimator behaves exactly as independent sampling predicts. The
excess risk falls faster than the feature error, so the readout does not merely inherit
feature noise but compounds it. The practical consequence is a mismatch in when each
quantity is good enough: the feature error reaches its $1\%$ tolerance near $10^4$
shots, while at that same budget the readout remains far above its $5\%$ band. Nothing
is wrong with the estimator; the shortfall is in how much feature precision least
squares demands before its output stops moving, and Appendix~\ref{app:forecasts} shows
what that looks like at the level of individual predictions.}

\subsection{\revc{Scaling with reservoir width and memory horizon}}
\label{sec:width}
\label{sec:decomposition}
\label{sec:horizon}

\begin{figure}[t]\centering
\includegraphics[width=0.95\textwidth]{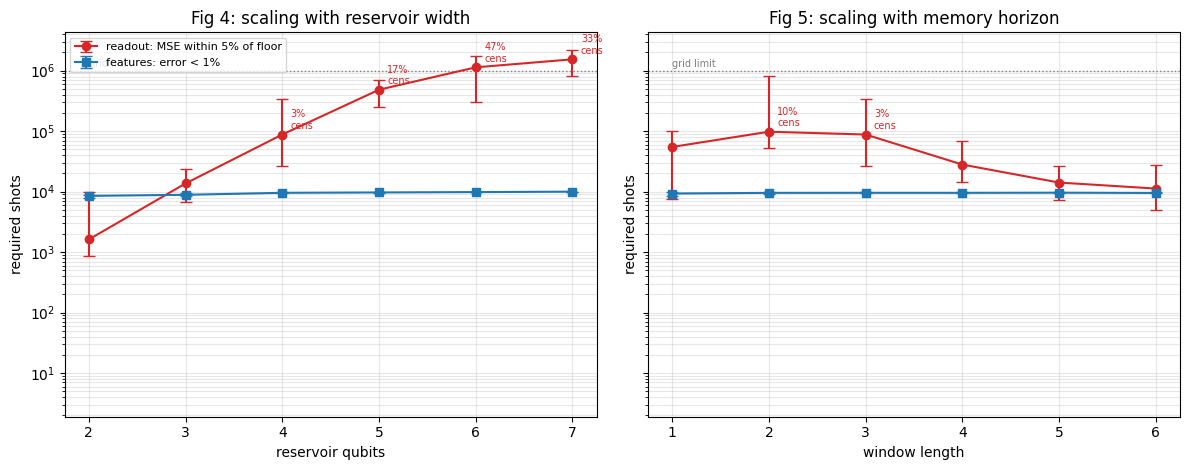}
\caption{Required shots against reservoir width (left) and memory horizon (right)
under both budgets. The precision budget (blue) is nearly flat; the convergence
budget (red) is not. Censoring fractions are annotated; heavily censored medians are
lower bounds.}
\label{fig:scaling}
\end{figure}

\revc{Width and window length are the two structural parameters of the reservoir, and
Fig.~\ref{fig:scaling} reports both under the same pair of budgets.}

\revc{\paragraph{Width.}}
As the reservoir widens from two to seven qubits the noise-free floor improves by
roughly $12\times$, the precision budget rises by only $17\%$, from \num{8511} to
\num{9998} shots, and the convergence budget grows by nearly three orders of
magnitude, from \num{1621} to \num{1541034} shots.

\revc{The precision budget barely moves because it is bounded. Each entry of
$X_N - X_\infty$ has variance $(1 - \langle Z_i\rangle_t^2)/N$, so
$\mathbb{E}[\varepsilon(N)^2] = (1 - \overline{\langle Z\rangle^2})/N$ with
$\overline{\langle Z\rangle^2}$ the mean of $\langle Z_i\rangle_t^2$ over all entries,
and}
\begin{equation}
N_\varepsilon \;=\; \frac{1 - \overline{\langle Z\rangle^2}}{\tau_{\mathrm{feat}}^2}
\;\le\; \tau_{\mathrm{feat}}^{-2} \;=\; 10^4 ,
\label{eq:neps}
\end{equation}
\revc{with equality when the expectation values vanish. Measured
$\overline{\langle Z\rangle^2}$ falls from $0.17$ at two qubits to $0.006$ at seven,
which is the concentration analysed in~\cite{sannia2025}, so $N_\varepsilon$ rises
toward its ceiling and flattens.
Equation~\eqref{eq:neps} reproduces every measured value to within $3.3\%$. A
power-law description would mislead: the quantity saturates rather than growing, so a
fitted exponent reflects the range fitted rather than any scaling.}

The growth of the convergence budget is far larger, but
Proposition~\ref{prop:confound} makes it uninterpretable on its own; a fixed-budget
comparison is used in Sec.~\ref{sec:decomposition} instead. Censoring is severe at the
upper end ($14/30$ seeds at $n_{\mathrm{res}} = 6$), so those medians are lower bounds
and the apparent flattening at large width is an artefact rather than a plateau.
\revc{Appendix~\ref{app:surface} shows the same data as a surface over budget and
size.}

\revc{\paragraph{Memory horizon.}}
\revc{The precision budget is likewise flat, \numrange{9312}{9612} shots throughout,
for the same reason.} The floor is not: it is flat for windows of one to three steps
and then degrades by nearly $4\times$ between window $3$ and window
$6$ (Appendix~\ref{app:floor}).

\revc{Depth is the tempting explanation, since by~\eqref{eq:depth} it grows linearly
in $k$. It is the wrong one.} These simulations are noiseless and the evolution is
unitary, so no information is lost; what changes is where it sits. Each additional
block moves earlier inputs further into higher-weight correlations while the readout
still measures only $n_{\mathrm{res}}$ single-body observables. This is a
\emph{readout-locality} limit: the constraint is the locality of the measured
operators, not the loss of information. \revc{Two facts favour it over depth. The
effect appears with no noise model at all, and enlarging the readout to two-body
operators recovers much of the loss without touching the circuit
(Sec.~\ref{sec:isobudget}) --- a depth-limited reservoir would not improve under a
change that leaves the circuit unchanged.}

\revc{On hardware the two coincide, since~\eqref{eq:depth} implies $2kn_{\mathrm{res}}$
entangling gates, so a reader meeting the window optimum there would reasonably assume
decoherence. In simulation it is not that.} The convergence budget falls at long
windows, from \num{97950} at $k=2$ to \num{11246} at $k=6$, which the
fixed-budget comparison below shows to be confounding rather than a genuine reduction
in measurement cost.

\revc{\paragraph{Comparing at a fixed budget.}}
Proposition~\ref{prop:confound} says that $N_\star$ mixes achievable performance with
noise sensitivity. The relative sampling penalty $\rho$ of~\eqref{eq:rho} separates
them without assuming a functional form, and under it the two sweeps come apart.
Across width, $\rho(10^4)$ grows monotonically and steeply: a budget costing a
two-qubit reservoir a few percent of its floor costs a five-qubit reservoir more than
its entire floor again, so the measurement cost of width is real and not merely the
consequence of a better target. Across memory horizon it does not. There
$N_\star$ varies by $8.7\times$ and non-monotonically, which read naively suggests
long windows are cheap, but Table~\ref{tab:horizon} shows the floor moving by nearly
$4\times$ over the same range in the opposite direction, which is sufficient
by~\eqref{eq:nstar} to produce that pattern with no change in noise sensitivity at
all. \revc{The two sweeps yield superficially similar $N_\star$ curves, both
non-monotonic and both spanning an order of magnitude, and they do not mean the same
thing. That contrast is why Proposition~\ref{prop:confound} is worth applying rather
than merely noting.}

\begin{table}[t]\centering
\caption{Scaling with reservoir width. ``Floor'' is the median analytic test MSE;
$N_\star$ and $N_\varepsilon$ are the convergence and precision
budgets; ``cens.'' is the number of seeds not reaching the convergence tolerance
within the grid.}
\label{tab:width}
\begin{tabular}{ccS[table-format=7.0]cS[table-format=5.0]}
\toprule
$n_{\mathrm{res}}$ & Floor & {$N_\star$} & cens. & {$N_\varepsilon$} \\
\midrule
2 & 0.0739 & 1621    & 0/30  & 8511 \\
3 & 0.0195 & 13668   & 0/30  & 8865 \\
4 & 0.0138 & 88021   & 1/30  & 9569 \\
5 & 0.0077 & 484693  & 5/30  & 9701 \\
6 & 0.0073 & 1136246 & 14/30 & 9835 \\
7 & 0.0062 & 1541034 & 10/30 & 9998 \\
\bottomrule
\end{tabular}
\end{table}
\begin{table}[t]\centering
\caption{Scaling with memory horizon (window length $k$). Columns as in
Table~\ref{tab:width}.}
\label{tab:horizon}
\begin{tabular}{ccS[table-format=6.0]cS[table-format=5.0]}
\toprule
$k$ & Floor & {$N_\star$} & cens. & {$N_\varepsilon$} \\
\midrule
1 & 0.0123 & 54830 & 0/30 & 9312 \\
2 & 0.0129 & 97950 & 3/30 & 9553 \\
3 & 0.0138 & 88021 & 1/30 & 9569 \\
4 & 0.0200 & 28029 & 0/30 & 9560 \\
5 & 0.0442 & 14140 & 0/30 & 9612 \\
6 & 0.0474 & 11246 & 0/30 & 9519 \\
\bottomrule
\end{tabular}
\end{table}

\subsection{\revc{Observable sets and task dependence}}
\label{sec:isobudget}
\label{sec:capacity}

\revc{Two choices remain that are not structural: which observables are read out, and
which task the features are used for.}

\revc{\paragraph{Observable sets.}}
Enlarging the readout without enlarging the reservoir is the strategy behind spatial
multiplexing~\cite{nakajima2019}. Four readout sets are compared: single-qubit $Z$;
$Z$ with two-qubit $ZZ$ correlators; $Z$ with $X$-basis features; and all three.
Correlators require no additional circuit executions, being computed from bitstrings
already measured, whereas $X$-basis features need a second execution, so at fixed
total budget the shots are split between bases rather than the richer set being given
more.

\begin{figure}[t]\centering
\includegraphics[width=0.95\textwidth]{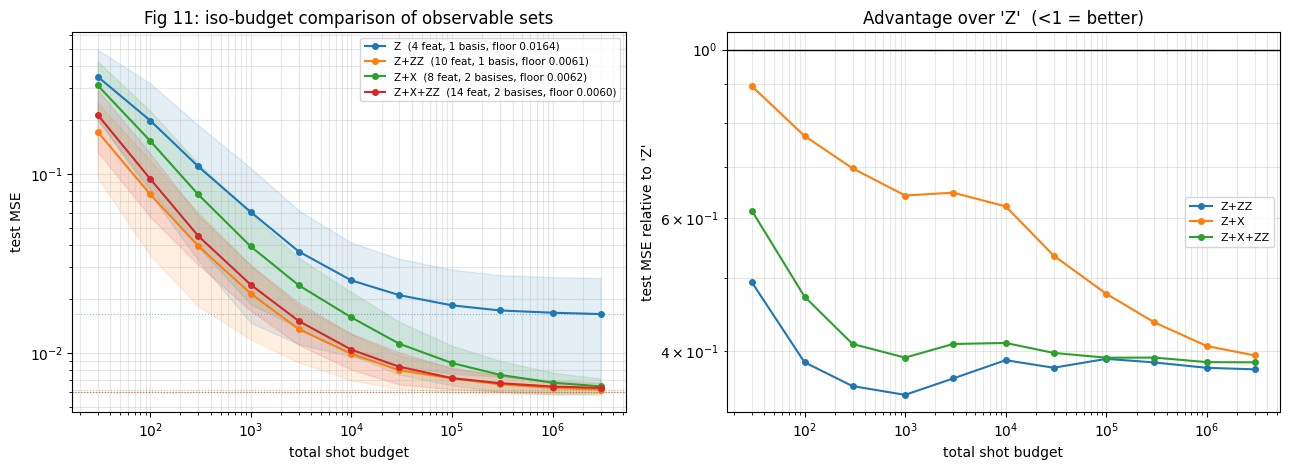}
\caption{Iso-budget comparison. Left: test MSE against total shot budget, with
noise-free floors dotted. Right: error relative to the $Z$ baseline; values below
unity indicate superiority. $Z + ZZ$ is below the baseline at every budget tested.}
\label{fig:isobudget}
\end{figure}

Table~\ref{tab:obs} shows the convergence budget ranking $Z$ cheapest by a wide
margin. That ranking is an artefact of Proposition~\ref{prop:confound}: each set is
graded against its own floor, and the floors differ by more than a factor of two.
Figure~\ref{fig:isobudget} compares the sets at equal total budget instead, which is
the comparison a fixed measurement allowance calls for. There $Z + ZZ$ is superior to $Z$ at every budget
tested, by a factor of \verifynum{$\approx\!2.3\times$} in test MSE, with no crossover
point. The $X$-basis set trails $Z + ZZ$ at small budgets, because splitting shots
across two bases costs more than the extra features return, and only becomes
competitive at large budgets; $Z + X + ZZ$ tracks $Z + ZZ$ closely, so $X$-basis
features add little once correlators are present. On these tasks, then, correlators computed from data already
measured were worth more than a second measurement basis. \revc{Appendix~\ref{app:ridge} reports a separate test of whether
ridge regularisation can substitute for measurement precision; it can, but only
marginally.}

\begin{figure}[t]\centering
\includegraphics[width=0.95\textwidth]{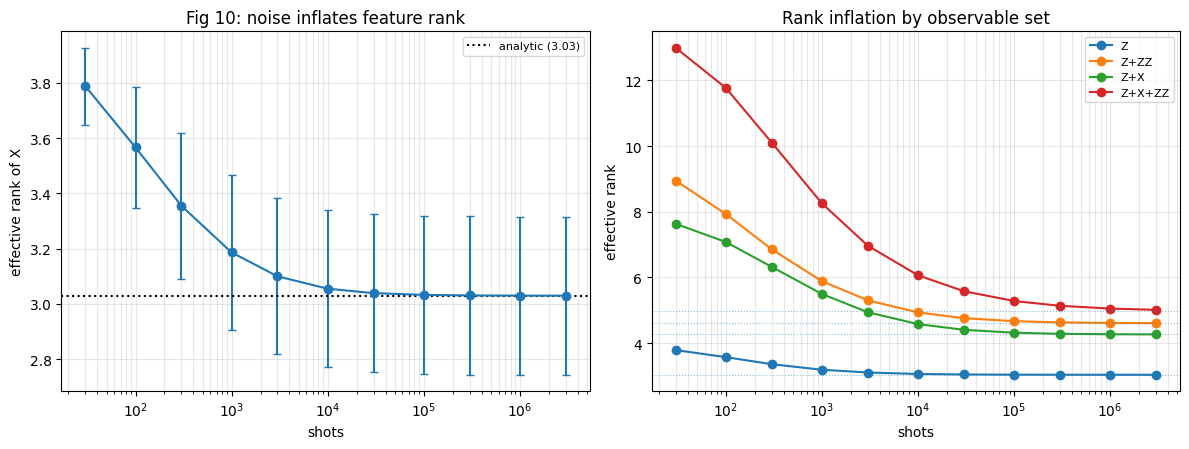}
\caption{Effective rank of the feature matrix against shot budget (left) and by
observable set (right). Sampling noise inflates rank; inflation is stronger for
larger feature sets.}
\label{fig:rank}
\end{figure}

\revc{Figure~\ref{fig:rank} shows one route by which this operates. Sampling noise supplies the feature
matrix with directions that carry no signal, and least squares fits whatever
directions it is given; the inflation of the matrix's effective rank measures how many
such directions are on offer, and relaxes to the noise-free value as the budget
grows.}

\begin{figure}[t]\centering
\includegraphics[width=0.95\textwidth]{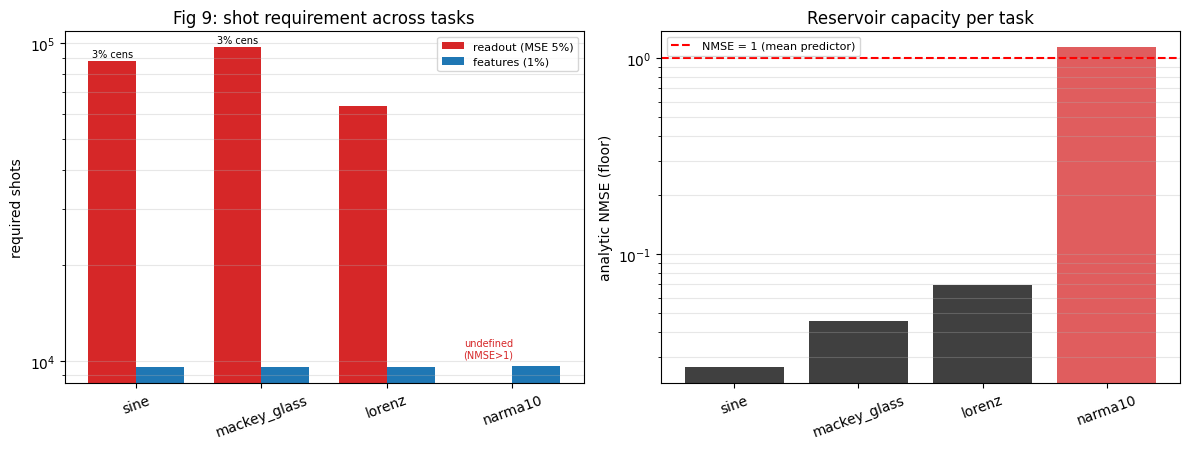}
\caption{Left: shot requirement by task under both criteria. Right: noise-free
normalised error; the dashed line marks the mean predictor. NARMA-10 lies above it.}
\label{fig:tasks}
\end{figure}

\revc{\paragraph{Task dependence.}}
The precision budget is invariant across tasks as well as configurations, and more
sharply so than across either structural sweep. Over the four tasks studied
(Table~\ref{tab:tasks}) it spans \numrange{9569}{9630} shots, a $0.6\%$ range, while
the noise-free error those tasks admit spans a factor of $43$, from
$\mathrm{NMSE} = 0.027$ on the sinusoid to $1.135$ on NARMA-10. \revc{This follows
from~\eqref{eq:neps}: the budget depends on the reservoir's expectation values, not on
what the features are later used for, including the case where they are of no use at
all.} The convergence budget, by contrast, again tracks the floor rather than any
intrinsic difficulty\revc{, as Fig.~\ref{fig:tasks} shows}.

\begin{table}[t]\centering
\caption{Observable sets. $N_\star$ ranks $Z$ cheapest, but each set is graded
against a different floor; see Fig.~\ref{fig:isobudget} for the iso-budget
comparison.}
\label{tab:obs}
\begin{tabular}{lccS[table-format=7.0]c}
\toprule
Observables & Features & Bases & {$N_\star$} & Floor \\
\midrule
$Z$          & 4  & 1 & 66864   & 0.0138 \\
$Z + ZZ$     & 10 & 1 & 290829  & 0.0061 \\
$Z + X$      & 8  & 2 & 923442  & 0.0061 \\
$Z + X + ZZ$ & 14 & 2 & 1857682 & 0.0060 \\
\bottomrule
\end{tabular}
\end{table}
\begin{table}[t]\centering
\caption{Task dependence at $n_{\mathrm{res}}=4$, $k=3$. The precision budget varies
by $0.6\%$ across tasks whose achievable error varies by a factor of $43$. NARMA-10
exceeds the mean predictor, so its convergence budget is undefined.}
\label{tab:tasks}
\begin{tabular}{lcS[table-format=6.0]cS[table-format=4.0]}
\toprule
Task & Floor (NMSE) & {$N_\star$} & cens. & {$N_\varepsilon$} \\
\midrule
Sinusoid      & 0.0266 & 88021 & 1/30 & 9569 \\
Mackey--Glass & 0.0455 & 97096 & 1/30 & 9575 \\
Lorenz-63     & 0.0698 & 63500 & 0/30 & 9572 \\
NARMA-10      & 1.1350 & {---} & {---} & 9630 \\
\bottomrule
\end{tabular}
\end{table}

\subsection{\revc{Classical baselines and capacity limits}}
\label{sec:baselines}

\revc{Measurement cost is the object of study here, not predictive performance, but a
reader is entitled to know whether the reservoir does anything a simpler model does
not. On these tasks it largely does not.}

The natural comparison is a model given exactly the same input: the delay window
$(u_{t-k+1},\dots,u_t)$ the reservoir receives. Three are used --- ordinary least
squares on the window, the same with per-element polynomial terms to cubic order, and
next-generation reservoir computing (NVAR), which augments the window with all
pairwise products of its elements~\cite{gauthier2021} --- all with identical splits,
washout and readout. On the autonomous tasks the window is nearly the answer, since
the target is the drive advanced one step, so linear regression is strong and on the
sinusoid beats the reservoir. On tasks built to require cross-delay products, such as
$y_t = u_t u_{t-\delta}$, linear and per-element-polynomial regression both fail at
normalised error $\approx 1$ while the reservoir with $Z + ZZ$ readout succeeds ---
the entangling dynamics generate cross-delay correlations and the two-body observables
expose them. \revc{But NVAR, which enumerates those products explicitly, solves the
same tasks essentially exactly, so the advantage holds only against baselines lacking
cross terms.}

\revc{NARMA-10 bounds the protocol in the other direction: it is not learnable by this
reservoir at any window length tested, with noise-free normalised error above unity at
$k = 3$ ($1.135$) and $k = 12$ ($1.252$), so its convergence budget is undefined and
excluded. Preliminary evidence attributes this to the protocol rather than to
capacity. Widening from $n_{\mathrm{res}} = 4$ to $8$ at $k = 12$ does not help
($1.23 \to 1.31$) while least squares on the identical window reaches $0.28$, so the
information is linearly available in the reservoir's own input and is lost in
encoding. The reservoir also degrades as the window lengthens while the classical
baseline improves, which is the readout-locality behaviour of
Sec.~\ref{sec:width} in sharper form. Appendix~\ref{app:narma} gives the diagnosis.}

\revc{This exposes a limitation in the task selection. The sinusoid, Lorenz and
Mackey--Glass tasks are all autonomous one-step-ahead problems differing in chaoticity
but not in problem class, and NARMA is the only driven task and beyond capacity. The
ladder therefore varies two things at once, and no difficulty-scaling claim is made.}

\section{Discussion}
\label{sec:discussion}

\paragraph{\revc{Which budget transfers to hardware.}}
\revc{The precision budget is a statement about estimating $\langle Z_i\rangle$ from
$N$ projective measurements, and Eq.~\eqref{eq:neps} fixes it in closed form, so it is
the quantity one would expect to carry over. The convergence budget is a property of
the least-squares estimator and its interaction with the noise-free floor, and is not
transferable as a shot budget at all; for comparing configurations the relative
sampling penalty at a stated budget,~\eqref{eq:rho}, is the appropriate quantity.}

\revc{That transfer is partial, and the limit is worth stating. On hardware the
estimated features are biased, not merely noisy: decoherence, gate infidelity and
readout error displace the expectation values, so no shot budget converges to the
analytic $X_\infty$ studied here. What these results establish is a sampling floor, to
which a hardware bias term must be added; no attempt is made to model that term.}

\paragraph{\revc{Relation to exponential concentration.}}
\revc{Equation~\eqref{eq:neps} makes the precision budget a function of
$\overline{\langle Z\rangle^2}$, which falls with width in these reservoirs. That is
the concentration analysed by Sannia \emph{et al.}~\cite{sannia2025}, and it also
drives the growth of the convergence budget: as expectation values crowd together,
resolving differences between them costs more. The sweeps here span
$n_{\mathrm{res}} \le 7$ and cannot distinguish exponential from polynomial growth,
so the functional form is left to a dedicated study.}

\paragraph{Reporting practice.}
The dominance of reservoir-seed variance (Sec.~\ref{sec:convergence}) implies that
single-instance QRC results are difficult to interpret. \revc{The seed component
exceeds shot noise by an order of magnitude or more at every budget, and the gap
widens as shots increase, so a result quoted for one reservoir draw is dominated by a
component unrelated to the dynamics under study. Seed ensembles with reported
dispersion would be a reasonable default in this literature.}

\paragraph{Design implications.}
\revc{Three observations bear on configuration choice in this setting. Correlators
computed from bitstrings already measured required no additional circuit executions
and outperformed the alternatives at every budget tested. Buffered history beyond a
few steps degraded performance through readout locality, which a richer observable set
only partly offset. And width carried a measurement cost that the precision budget
does not reveal: at $10^4$ shots the relative sampling penalty rose from a few percent
of the floor at two qubits to more than the floor itself at five.}

\revc{These are observations about one circuit family at small scale. The confounding
of Proposition~\ref{prop:confound}, by contrast, follows from the definition of
$N_\star$ and does not depend on the reservoir studied.}

\section{Conclusion}

\revc{This paper has characterised the measurement cost of a small restart-protocol
quantum reservoir in simulation, separating two quantities that the phrase ``required
shots'' conflates. The precision budget, the shots needed to estimate the reservoir
features to fixed relative accuracy, admits the closed form~\eqref{eq:neps} and is
bounded above by $\tau_{\mathrm{feat}}^{-2}$; for these reservoirs it sits near
$10^4$ shots and varies by under $20\%$ across width, under $4\%$ across memory
horizon, and under $1\%$ across four tasks whose achievable error spans a factor of
$43$. The convergence budget, the shots needed for the trained readout to approach its
own noise-free ceiling, varies by three orders of magnitude over the same sweeps, but
Proposition~\ref{prop:confound} shows that it depends on that ceiling by construction
and cannot be read as measurement cost. Compared at a fixed budget, width carries a
real and steep cost while window length does not.}

\revc{Three further results bear on practice. Reservoir-seed variance exceeds shot
noise by one to two orders of magnitude at every budget. Buffered memory has a short
optimum, attributable to the locality of the measured operators rather than to circuit
depth. And two-qubit correlators, obtainable from bitstrings already measured,
outperform single-qubit features at every budget tested.}

\revc{The scope is narrow. This is one circuit family of at most seven qubits, in
simulation, on tasks that are predominantly autonomous one-step-ahead problems, and
classical regression on the same delay window remains competitive throughout --- no
advantage claim is made or supported. The excess-risk exponent drifts with width in a
way this work does not explain, and on hardware a bias term would sit above the
sampling floor measured here.}

\revc{The natural extensions follow from those limits. More realistic datasets, driven
rather than autonomous and drawn from applications rather than synthetic benchmarks,
would test whether the task invariance reported here survives contact with data whose
structure the reservoir was not implicitly handed. Other classes of quantum reservoir,
with different topologies, encodings and entangling structure, would establish which
of these findings are properties of QRC and which are properties of the star-topology
circuit studied. Equation~\eqref{eq:neps} suggests where to look first: the precision
budget is governed by how strongly a reservoir concentrates its expectation values, so
reservoirs that concentrate less should be measurably cheaper to read.}

\section*{Acknowledgments}
This work was supported by Allstate. The author is sincerely grateful for the support of Prof. Prem Kumar (Northwestern University), which made this work possible.
This research was supported in part through the computational resources and staff contributions provided for the Quest high performance computing facility at Northwestern University which is jointly supported by the Office of the Provost, the Office for Research, and Northwestern University Information Technology.

\clearpage
\appendix

\FloatBarrier
\section{\revc{Regularisation}}
\label{app:ridge}
\label{sec:ridge}

\begin{figure}[h]\centering
\includegraphics[width=0.95\textwidth]{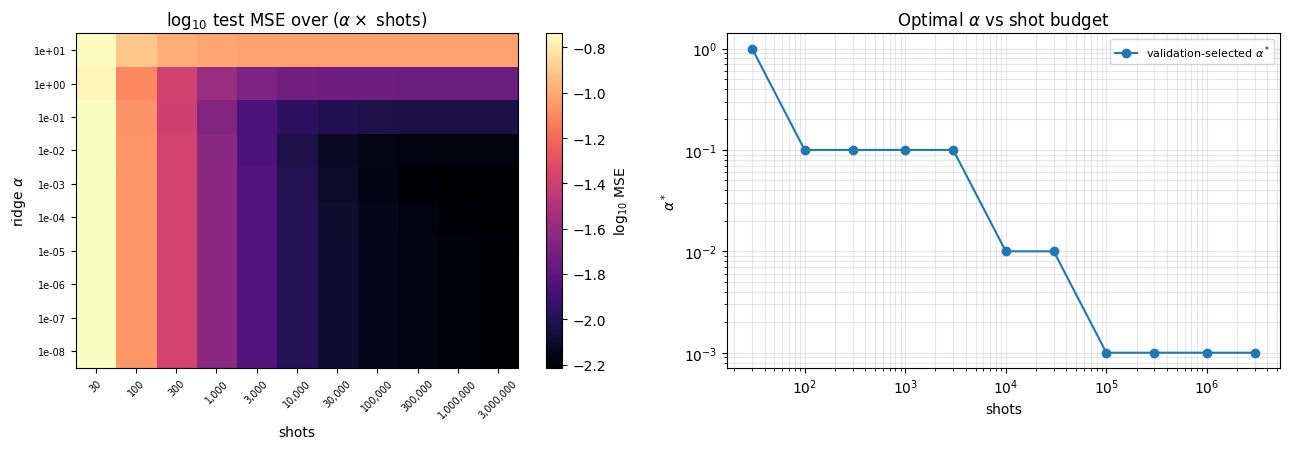}
\caption{Left: test MSE over the (penalty $\times$ budget) grid. Right:
validation-selected penalty against budget, falling as sampling noise decreases.}
\label{fig:ridge}
\end{figure}

\revc{Whether ridge regularisation can substitute for measurement precision was tested
separately. With the four-dimensional $Z$ feature set and $\sim\!95$ training points
the question is ill-posed, since there is little to regularise, and the
validation-selected penalty sits at the grid floor for every budget. With the
fourteen-dimensional $Z + X + ZZ$ set it becomes testable and is weakly confirmed: the
selected penalty falls by two orders of magnitude as the budget grows, from
\verifynum{$\alpha^\star \approx 10^{-1}$} at $10^2$ shots to
\verifynum{$\approx\!10^{-3}$} at $10^5$, and ridge improves on ordinary least squares
by \verifynum{$2$--$4\%$}, the advantage vanishing in the noise-free limit.
Regularisation therefore compensates for sampling noise only partially, and by far
less than the choice of observable set. Figure~\ref{fig:ridge} shows the penalty
surface and the selected penalty against budget.}

\FloatBarrier
\section{\revc{Error surface over budget and size}}
\label{app:surface}

\begin{figure}[h]\centering
\includegraphics[width=0.95\textwidth]{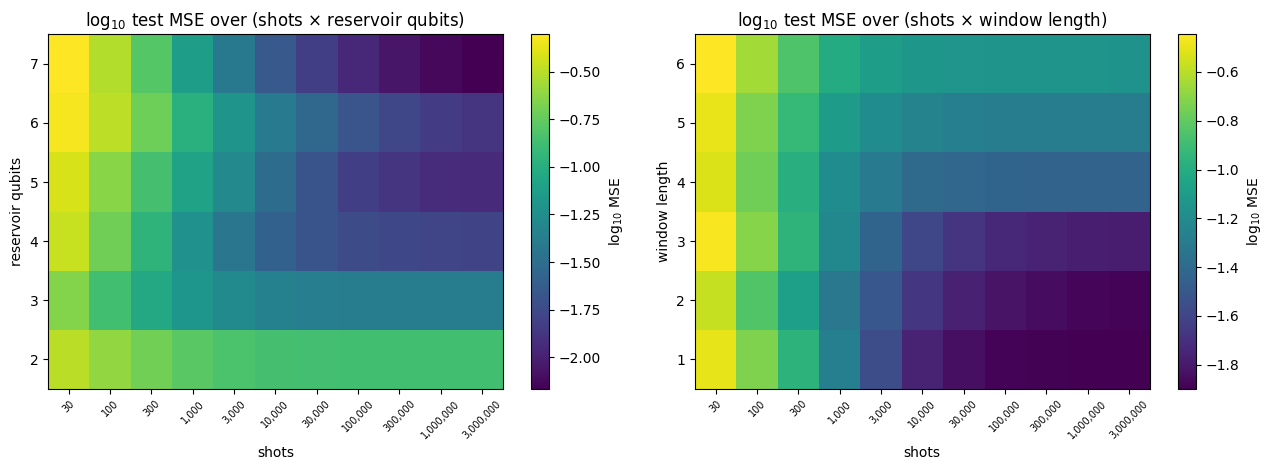}
\caption{Test error over the (budget $\times$ width) and (budget $\times$ window)
surfaces. Contours running diagonally would indicate that budget and size trade
against each other; the near-vertical contours in the right panel instead show
window length and budget acting almost independently.}
\label{fig:surface}
\end{figure}

\revc{Figure~\ref{fig:surface} presents the width and horizon sweeps of
Sec.~\ref{sec:width} as surfaces over shot budget and structural parameter, which
shows at a glance whether budget and size trade against one another.}

\FloatBarrier
\section{\revc{NARMA-10 diagnosis}}
\label{app:narma}

\revc{NARMA-10 requires the product $u(t-9)\,u(t)$ together with an accumulating
autoregressive target, and the reservoir sees only the exogenous drive. The table
below separates three candidate explanations for its failure: that the task is
unlearnable from the window supplied, that the reservoir lacks capacity, or that the
encoding loses information the window contains. Ordinary least squares on the
identical window reaches $\mathrm{NMSE} = 0.28$ at $k = 12$, so the first is
excluded; widening the reservoir from four to eight qubits does not help, so the
second is excluded; and the reservoir degrades as the window lengthens while the
classical baseline improves, which leaves the third. Table~\ref{tab:narma} gives the figures. The run used four
reservoir seeds and should be repeated at production scale before being relied upon.}

\begin{table}[h]\centering
\caption{\revc{NARMA-10 normalised error. Classical models improve with window length;
the reservoir does not, and widening it does not help.}}
\label{tab:narma}
\begin{tabular}{cccccc}
\toprule
$k$ & OLS-linear & OLS-cubic & QRC $n{=}4$ & QRC $n{=}6$ & QRC $n{=}8$ \\
\midrule
3  & 0.971 & 1.016 & 1.120 & 1.104 & 1.090 \\
6  & 0.839 & 1.037 & 1.199 & 1.194 & 1.166 \\
12 & 0.280 & 0.310 & 1.228 & 1.197 & 1.308 \\
\bottomrule
\end{tabular}
\end{table}

\FloatBarrier
\section{\revc{Noise-free floor against width and horizon}}
\label{app:floor}

\begin{figure}[h]\centering
\includegraphics[width=0.88\textwidth]{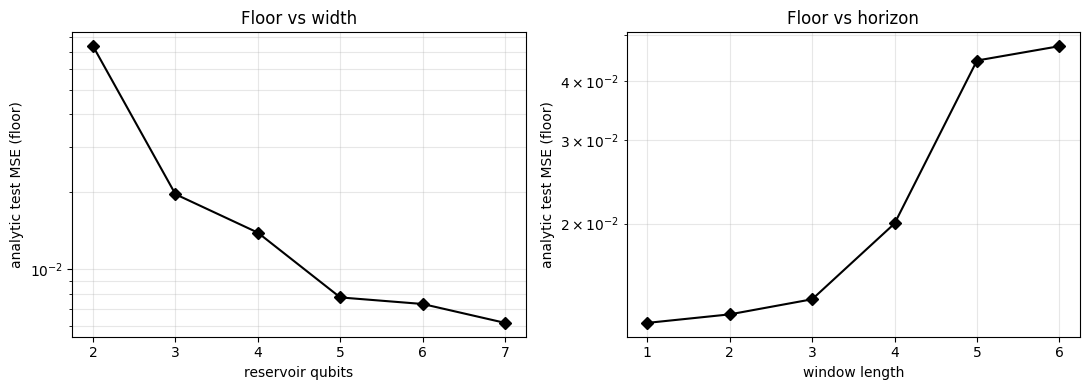}
\caption{Noise-free error floor against reservoir width (left) and memory horizon
(right). This panel is necessary to read Fig.~\ref{fig:scaling}: by
Proposition~\ref{prop:confound} the convergence budget moves inversely with the
floor, so the fall in $N_\star$ at long windows reflects the floor degrading rather
than measurement becoming cheaper.}
\label{fig:floor}
\end{figure}

\revc{Figure~\ref{fig:floor} plots the noise-free floor directly for both structural
sweeps. The same values appear in the ``Floor'' columns of Tables~\ref{tab:width}
and~\ref{tab:horizon}; the figure makes the non-monotonicity in window length easier
to see.}

\FloatBarrier
\section{\revc{Variance components as a share of the total}}
\label{app:variance}

\begin{figure}[h]\centering
\includegraphics[width=0.62\textwidth]{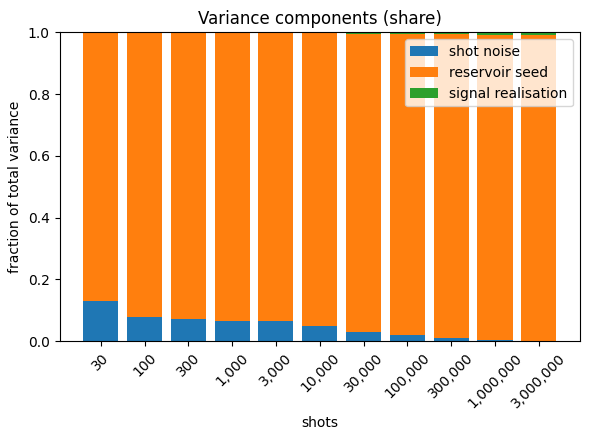}
\caption{\revc{Variance of the test error by component, as a share of the total at each
shot budget.}}
\label{fig:varshare}
\end{figure}

\revc{Figure~\ref{fig:varshare} gives the decomposition of
Sec.~\ref{sec:convergence} as a fraction of total variance at each budget, which makes
the dominance of the reservoir-seed component easier to read off than the absolute
scale does.}

\FloatBarrier
\section{Forecasts at representative shot budgets}
\label{app:forecasts}

\begin{figure}[h]\centering
\includegraphics[width=0.84\textwidth]{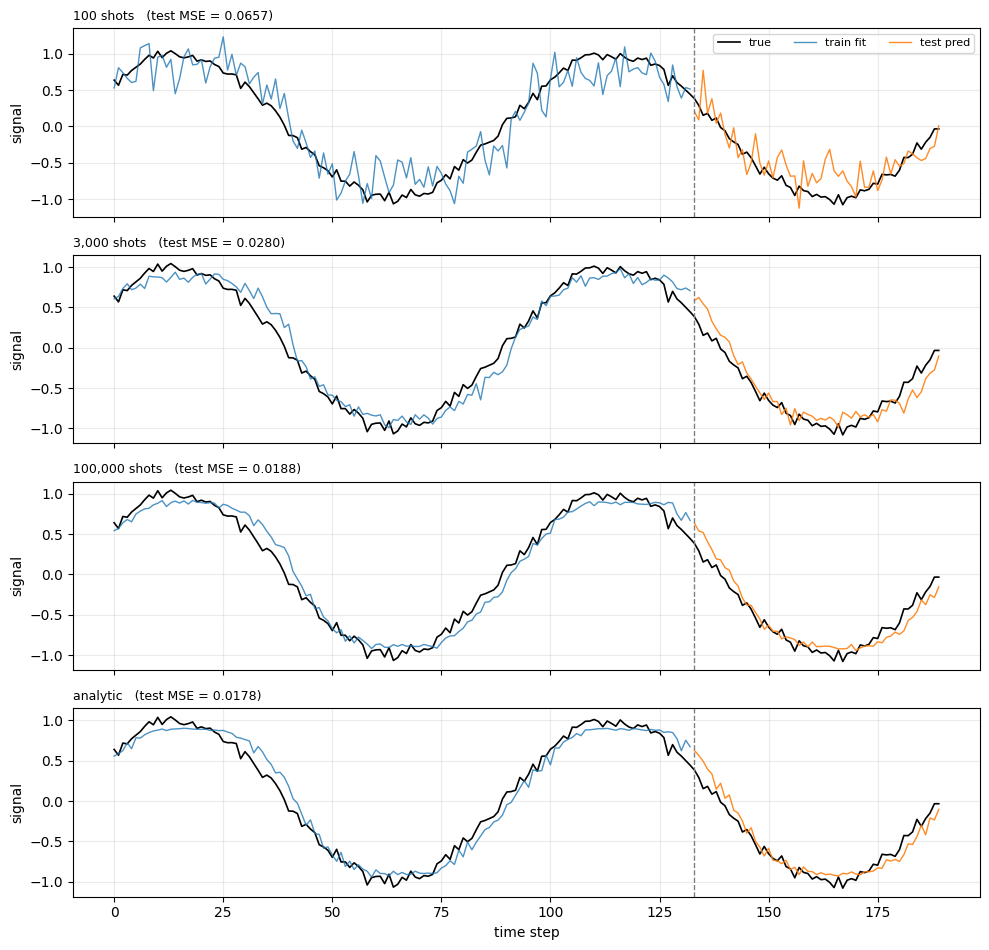}
\caption{One-step-ahead forecasts at three budgets and in the noise-free limit.
Under-sampled features produce a readout that tracks the signal's gross shape while
fitting sampling noise in detail; the failure is not a loss of the trend but a loss
of precision around it.}
\label{fig:forecasts}
\end{figure}

Figure~\ref{fig:forecasts} shows the one-step-ahead forecast on the sinusoid at three
shot budgets and in the noise-free limit, for the base configuration
($n_{\mathrm{res}} = 4$, $k = 3$).

\end{document}